\documentclass[a4paper,11pt]{article}
\usepackage{pos}

\title{Prospects to scrutinise or smash SM*A*S*H}

\author*[a]{Andreas Ringwald}

\affiliation[a]{Deutsches Elektronen-Synchrotron DESY,\\
  Notkestr. 85, 22607 Hamburg, Germany}

\emailAdd{andreas.ringwald@desy.de}

\abstract{
SM*A*S*H is an extension of the Standard Model of particle physics 
which has just the minimal number of fields in order to solve six puzzles 
of particle physics and cosmology in one smash: 
vacuum stability, inflation, baryon asymmetry, neutrino masses, strong CP, and dark matter. 
The parameters of SM*A*S*H are constrained by symmetries and 
requirements to solve these puzzles.
This provides various firm predictions for observables which can be confronted with experiments. 
We discuss the prospects and timeline to scrutinise or smash SM*A*S*H by cosmic microwave background polarisation experiments, 
axion haloscopes, and future space-borne gravitational wave detectors. 
}

\FullConference{1st General Meeting and 1st Training School of the COST Action COSMIC WISPers (COSMICWISPers)\\ 
5-14 September, 2023\\
Bari and Lecce, Italy\\}

\begin{document}
\maketitle

\section{SM*A*S*H}

The acronym ``SM*A*S*H'' stands for ``Standard Model*Axion*Seesaw*Higgs-Portal Inflation'' 
-- a minimal extension of the Standard Model (SM) which solves six puzzles 
of particle physics and cosmology in one smash~\cite{Ballesteros:2016euj,Ballesteros:2016xej}: 
vacuum stability, inflation, baryon asymmetry, neutrino masses, strong CP, and dark matter.\footnote{Similar models
have been considered in Refs.~\cite{Boucenna:2014uma,Ballesteros:2019tvf,Sopov:2022bog,Berbig:2022pye,Dutta:2023lbw}.}

\begin{table}[h]
\begin{center}
$\begin{array}{|c|c|c|c|c|c|c|c|c|}
\hline
  q & u & d & L & N & E    & Q &\tilde Q & \sigma  \\
\hline
 1/2 & -1/2 & -1/2 & 1/2 & -1/2 & -1/2   & -1/2 & -1/2 &1 
\\[.5ex]
\hline
\end{array}$
\end{center}
\vskip-.3cm
\caption{\small PQ-charge assignments of the fields in SM*A*S*H. 
The remaining SM fields have no PQ charge. 
\label{tab:pq_charges}}
\end{table}

A SM-singlet complex scalar field  $\sigma$ (the Peccei-Quinn (PQ) field),
a vector-like quark $Q$ and three SM-singlet neutrinos $N_i$, with $i=1,2,3$, are added to the SM. All the new fields, as well as the quarks and leptons of the SM, are assumed to be charged under a global $U(1)_{\rm PQ}$ symmetry, cf. Table~\ref{tab:pq_charges}.
The scalar potential respecting this PQ symmetry has the general form: 
$V(H,\sigma )= \lambda_H \left( H^\dagger H - \frac{v^2}{2}\right)^2
+\lambda_\sigma \left( |\sigma |^2 - \frac{v_{\sigma}^2}{2}\right)^2
+
2\lambda_{H\sigma} \left( H^\dagger H - \frac{v^2}{2}\right) \left( |\sigma |^2 - \frac{v_{\sigma}^2}{2}\right)$,
where $H$ is the SM Higgs doublet. In order to ensure that both the PQ and the electroweak symmetry are broken by the 
vacuum expectation values $\langle H^\dagger H\rangle = v^2/2$, $\langle |\sigma |^2\rangle=v_{\sigma}^2/2$,
where $v_\sigma\gg v \simeq 246$\,GeV, the scalar couplings are required to obey $\lambda_H, \lambda_\sigma >0$,  
$\lambda_{H\sigma}^2 <  \lambda_H \lambda_\sigma$. 
The $U(1)_Y$ hypercharge of the vector-like quark $Q$ is required to be 
$-1/3$, such that the most general Yukawa interactions of $Q$ and $N_i$, allowed by SM gauge and PQ symmetries, are  
${\cal L}\supset 
-[F_{ij}\bar{ N}_j P_L L_i\epsilon H+\frac{1}{2}Y_{ij}\sigma \bar N_i P_L  N_j 
+y\, \sigma \bar Q P_L Q+\,{y_{Q_d}}_{i}\sigma\bar{D}_iP_L Q +h.c.]$, 
where  $D_i$, $L_i$ denote the Dirac spinors associated with the down quarks and leptons of the $i$th generation, 
while the $N_i$ are taken to be Majorana spinors.
Electroweak vacuum instability~\cite{Degrassi:2012ry} -- the instability of the Higgs potential at large field values, present for the preferred value of the top mass -- can be avoided in SM*A*S*H by the stabilizing effect of the Higgs portal coupling 
$\lambda_{H\sigma}$~\cite{Lebedev:2012zw,Elias-Miro:2012eoi}. 
This requires $\lambda^2_{H\sigma}/\lambda_\sigma$ to be between $\sim 10^{-2}$ and $\sim10^{-1}$~\cite{Ballesteros:2016xej}. 
Inflation is realised in SM*A*S*H via non-minimal chaotic inflation.  
The inflaton is comprised by a mixture of the real part of the PQ field and the modulus of the Higgs. 
Reheating of the Universe after inflation proceeds then efficiently via the Higgs portal. 
SM*A*S*H predicts a lower bound on the ratio of the power in tensor to scalar fluctuations, $r \gtrsim 0.004$~\cite{Ballesteros:2016euj,Ballesteros:2016xej}, 
 a reheating temperature around $10^{12}$\,GeV~\cite{Ringwald:2022xif}, and a second order PQ phase transition at 
around $T_c \sim 10^8$\,GeV~\cite{Ballesteros:2016euj,Ballesteros:2016xej}.
The strong CP puzzle is solved in SM*A*S*H by the PQ mechanism~\cite{Peccei:1977hh}. 
The axion~\cite{Weinberg:1977ma,Wilczek:1977pj} -- the pseudo Goldstone boson associated with the spontaneous breaking of the PQ symmetry --   is the prime candidate for cold dark matter in SM*A*S*H~\cite{Preskill:1982cy,Abbott:1982af,Dine:1982ah}. Requiring the latter to constitute 100\% of dark matter, its decay constant,  
$f_a = v_\sigma$, is predicted to be in the range
$10^{10}\,{\rm GeV} \lesssim f_a <  2\times 10^{11}\,\mathrm{GeV}$.
The PQ symmetry breaking scale also gives rise to large Majorana masses for the heavy neutrinos. This can explain the smallness of the masses of the active neutrions through the seesaw mechanism~\cite{Minkowski:1977sc,Gell-Mann:1979vob,Yanagida:1979as,Mohapatra:1979ia} and results in the generation of the baryon asymmetry of the universe via thermal leptogenesis~\cite{Fukugita:1986hr}. 

In summary: The parameters in SM*A*S*H are constrained by symmetries and 
requirements to solve the puzzles of particle physics and cosmology, respectively. 
This provides various firm predictions of observables which can be confronted with experiments. 
In this contribution I will focus on some of these predictions. The order of discussion follows the approximate timeline of  
experiments being able to probe the predictions from SM*A*S*H.

\begin{figure}[t]
\begin{center}
\includegraphics[width=.32\textwidth]{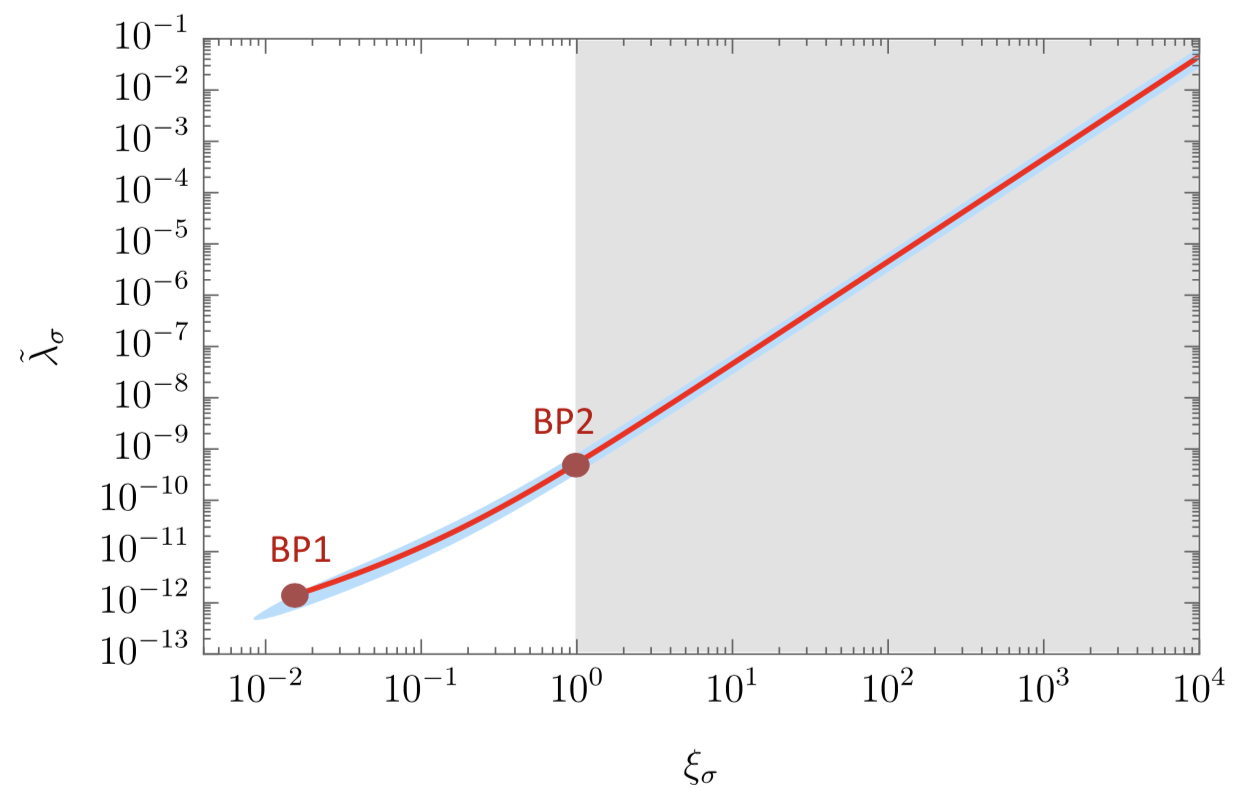}
\includegraphics[width=.32\textwidth]{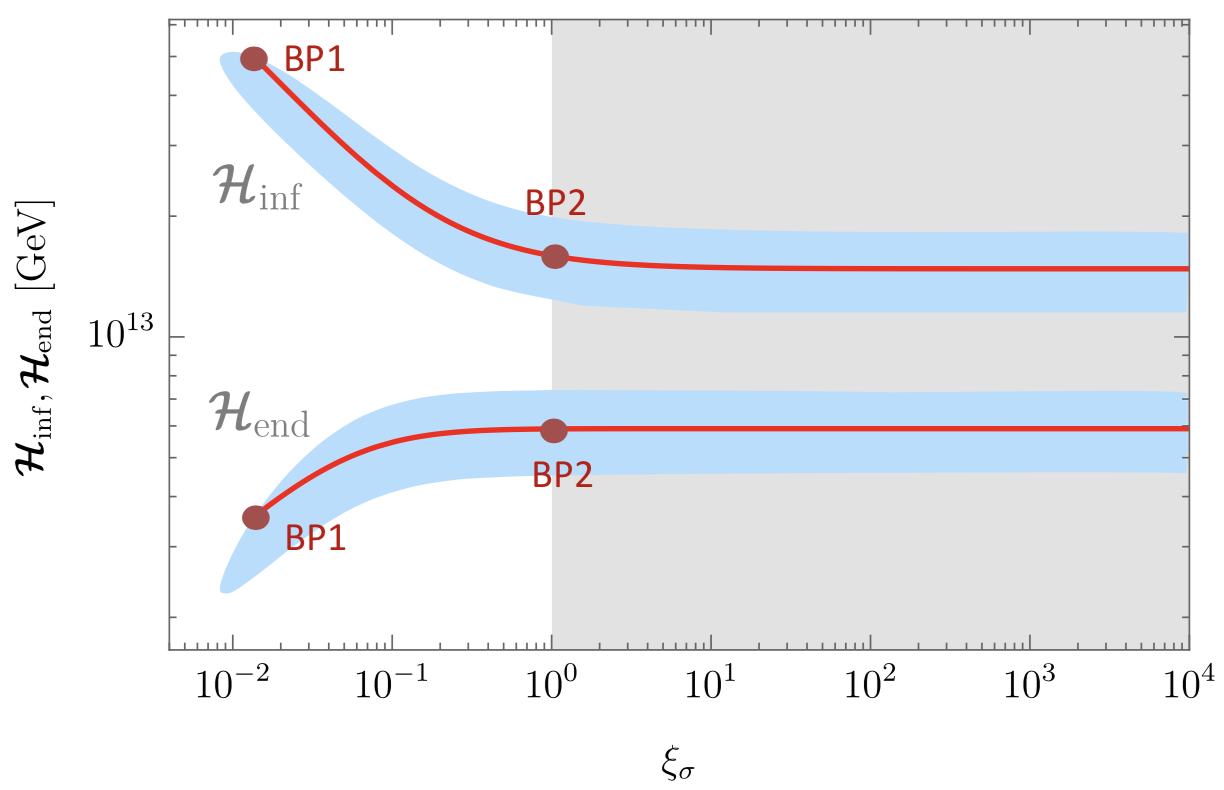}
\includegraphics[width=.32\textwidth]{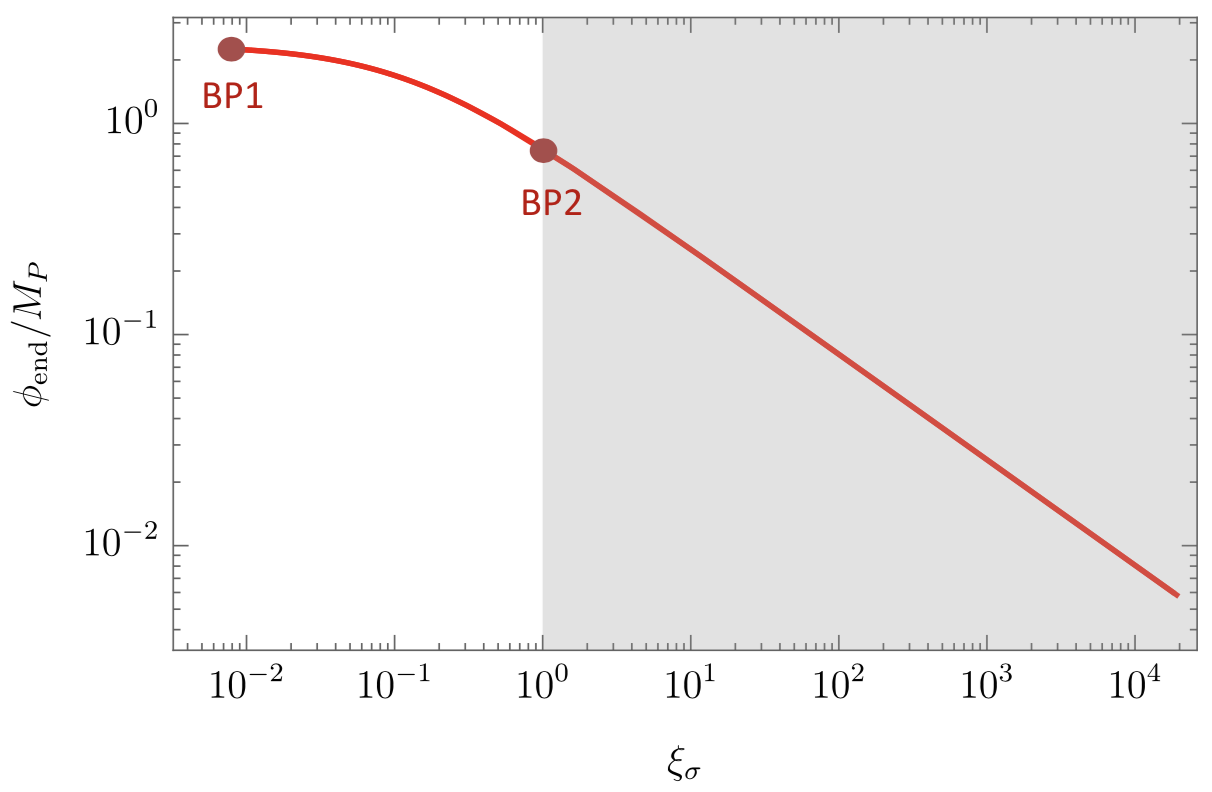}
\vskip-.2cm 
\caption{\label{fig:inflation_constraints}{Inflationary constraints/predictions on $\tilde\lambda_\sigma$ (left), Hubble scale at the beginning and end of inflation (center) and field value at the end of inflation (right) as a function of $\xi_\sigma$ and for $k_*=0.002$\,Mpc$^{-1}$~\cite{Ringwald:2022xif}. 
The blue regions are compatible within 95\% C.L. with the  latest  combination of Planck and BICEP/KECK data~\cite{BICEP:2021xfz}.
The red lines correspond to the predictions when taking into account the SM*A*S*H prediction of radiation domination immediately after inflation, and the red dots correspond to the benchmark scenarios BP1 and BP2. The gray shaded regions to the right of $\xi_\sigma>1$ indicate the region where the predictions may be threatened by the breakdown of peturbative unitarity. Figures adapted from Ref.~\cite{Ringwald:2022xif}.
}}
\end{center}
\end{figure}

\section{Confronting SM*A*S*H with Cosmic Microwave Background Observations} 

\begin{figure}[t]
\begin{center}
\includegraphics[width=.9\textwidth]{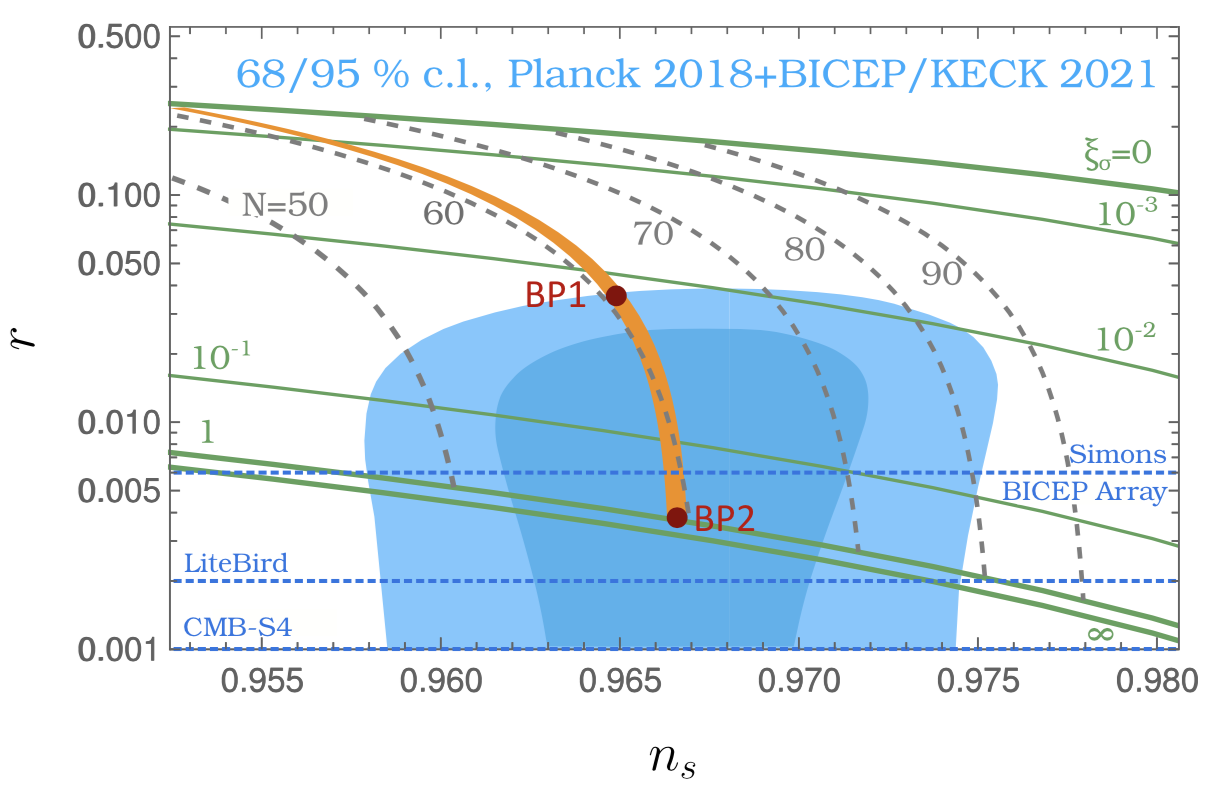}
\vskip-.2cm 
\caption{\label{fig:r_vs_ns} 
Inflationary predictions in SM*A*S*H
in the $r$ vs $n_{s}$ plane for $k_*=0.002$\,Mpc$^{-1}$.
The green solid/dashed gray lines are contours of constant 
$\xi_\sigma$/number of efolds, respectively. Accounting for the post-inflationary expansion history gives the orange region, and the red dots correspond to the benchmark scenarios BP1 and BP2. We also show the 68\% and 95\% C.L. contours arising from Planck and BICEP/KECK data \cite{BICEP:2021xfz}, as well as the 95\% projected sensitivities from BICEP Array~\cite{BICEP:2021xfz}, the Simons observatory~\cite{SimonsObservatory:2018koc}, LiteBird~\cite{LiteBIRD:2022cnt},  and  CMB-S4~\cite{Abazajian:2019eic}.  Figure adapted from Ref.~\cite{Ringwald:2022xif}.}
\end{center}
\end{figure}

\begin{figure}[h!]
\begin{center}
\includegraphics[width=0.95\textwidth]{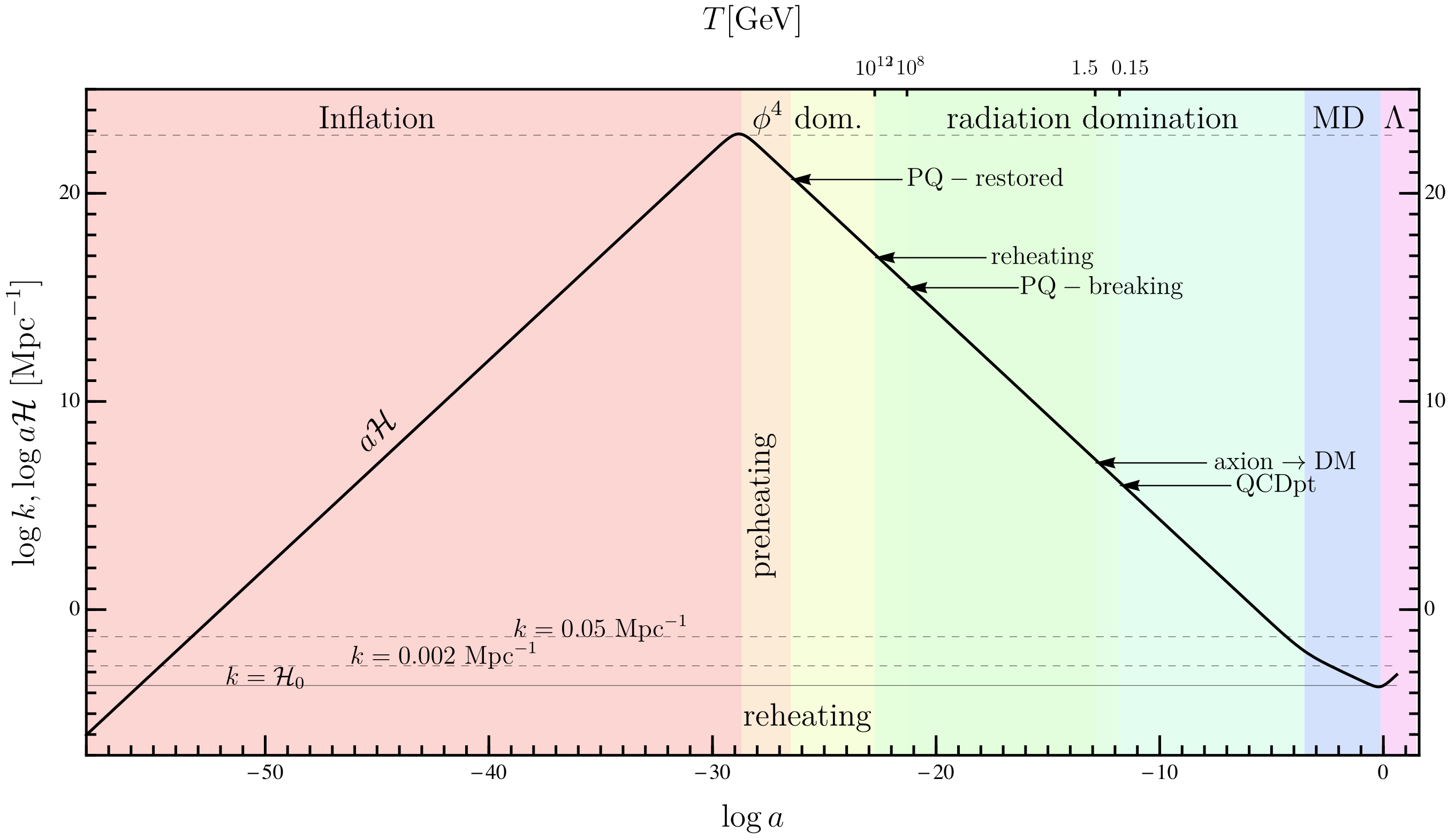}
\caption{\small The expansion and thermal history of the Universe in SM*A*S*H, emphasising the transition from inflation to radiation-domination-like expansion $a{\cal H}\propto 1/a$ before standard matter and cosmological constant domination epochs. 
Figure adapted from Ref.~\cite{Ballesteros:2016xej}.}  
\label{fig:expansion_history_smash}       
\end{center}
\end{figure}

Inflation is realised in SM*A*S*H by the dynamics of the PQ  and Higgs fields in the presence of their generically present non-minimal gravitational couplings 
to the Ricci scalar $R$~\cite{Spokoiny:1984bd,Futamase:1987ua,Salopek:1988qh,Fakir:1990eg,Bezrukov:2007ep}, described, 
in the Jordan frame,  by the action 
$S\supset - \int d^4x\sqrt{- g}\,\left[
     \frac{M^2}{2}  + \xi_H\, H^\dagger H+\xi_\sigma\, \sigma^* \sigma  
  \right] R$.
The mass scale $M$ is related to the reduced Planck mass ($M_P\simeq 2.435\times 10^{18}\,\rm GeV$) by  
$M^2_P=M^2+\xi_H v^2+\xi_\sigma v^2_\sigma$. 
After a Weyl transformation of the metric to the Einstein frame, the scalar potential  becomes flat for large field values.
Eventual problems with perturbative unitarity~\cite{Barbon:2009ya,Burgess:2009ea} are avoided by requiring $1\gtrsim \xi_\sigma\gg \xi_H\geq 0$. 
For positive $\lambda_{H\sigma}$, inflation can take place along the direction of $\phi\equiv \sqrt{2}{\rm Re}\,\sigma$, but in this case reheating can be shown to be problematic, leading to an excess of dark radiation~\cite{Ballesteros:2016euj,Ballesteros:2016xej}. 
In the case of a negative Higgs portal coupling,  $\lambda_{H\sigma}<0$, slow-roll inflation takes place along a valley in the scalar potential 
that can be approximated by the line $h/\phi=\sqrt{-\lambda_{H\sigma}/\lambda_H}$,  where $h$ is the neutral component of the Higgs doublet in the unitary gauge.  
The potential along the inflaton valley, $\tilde V(\chi) = \frac{1}{4} \tilde \lambda_\sigma
\phi(\chi)^4\left(1+\xi_{\sigma}\frac{\phi(\chi)^2}{M_P^2}\right)^{-2}$, is determined by two model parameters: an effective quartic coupling $\tilde\lambda_\sigma=\lambda_\sigma-\lambda_{H\sigma}^2/\lambda_H$ and the non-minimal coupling $\xi_\sigma$.
Here, $\phi$ and the canonically normalized inflaton field $\chi$ are related by
$\Omega^2\,d\chi/d\phi\simeq (b\,\Omega^2+6\,\xi_\sigma^2\,\phi^2/M_P^2)^{1/2}$, 
with $\Omega^2= 1+\xi_{\sigma}\frac{\phi(\chi)^2}{M_P^2}$ and $b=1+|\lambda_{H\sigma}/\lambda_H|$.
Quantum fluctuations during slow-roll inflation along this potential produce power spectra of 
scalar metric perturbations (density waves)
and 
tensor metric perturbations (gravitational waves (GWs)) 
which can be parametrized as 
$
\Delta^2_{s/t}(k)=A_{s/t}(k_*)\left({k}/{k_*}\right)^{n_{s/t}(k_*)-1+\cdots}$, 
where $k_*$ is a given reference pivot scale. 
Fitting the amplitude of these perturbations inferred from the observed cosmic microwave background (CMB) temperature and polarisation maps imposes one relation between these two parameters $\tilde\lambda_\sigma$, $\xi_\sigma$, which is approximately given by 
$7\times 10^{-3} \lesssim \xi_\sigma \simeq 4\times 10^{4}\sqrt{\tilde\lambda_\sigma}\lesssim 1$, see Fig.~\ref{fig:inflation_constraints} (left).
Accordingly, quantities during inflation can be characterized by a single parameter, which can be chosen as $\xi_\sigma$, as illustrated in Fig.~\ref{fig:inflation_constraints} (middle and right). 
The numerical predictions in the $r$ vs $n_s$ plane 
are shown in Fig.~\ref{fig:r_vs_ns} for $k_*=0.002$\,Mpc$^{-1}$. 
The thin solid lines indicate the values of the non-minimal coupling,  bounded by the quartic chaotic inflation limit, $\xi_\sigma=0$, and the $\xi_\sigma\rightarrow \infty$ limit. The dashed lines show the number of e-folds, $N=\Delta \log a$, of the cosmic scale factor $a$
between the pivot scale's crossing of the horizon and the end of inflation, cf. Fig.~\ref{fig:expansion_history_smash}. 
The figure also shows the 68\% and 95\% C.L.  regions from the combined data analysis of Planck and 
BICEP/KECK~\cite{BICEP:2021xfz}  at $k_*=0.002$\,Mpc$^{-1}$. 
We see, that $\xi_\sigma > 10^{-2}$ is perfectly compatible with current bounds on $n_s$ and $r$ and it also gives an adequate $N$. 
%

The latter can be determined by using the post-inflationary evolution of the universe  to match the scales of current CMB perturbations to their values at horizon crossing during inflation. 
Inflation in SM*A*S*H ends when $\phi\sim \mathcal{O}(M_P)$, see Fig.~\ref{fig:inflation_constraints} (right), after which the background goes through Hubble-damped oscillations that have the equation of state of a radiation bath. Hence radiation domination starts immediately after inflation, which fixes $N$ to around $60$, as can be inferred from 
Fig.~\ref{fig:expansion_history_smash}.  
This narrows the SM*A*S*H prediction to the thick orange line in Fig.~\ref{fig:r_vs_ns}.  
This prediction will be  probed exhaustively in the next decade by CMB polarisation experiments, such as BICEP Array~\cite{BICEP:2021xfz}, the Simons observatory~\cite{SimonsObservatory:2018koc}, LiteBird~\cite{LiteBIRD:2022cnt},  and  CMB-S4~\cite{Abazajian:2019eic}, see Fig.~\ref{fig:r_vs_ns}.
We conclude:
{\bf SM*A*S*H is smashed if CMB-S4 or LiteBird do not discover B modes from inflationary tensor metric perturbations (GWs)!}

\section{Scrutinising SM*A*S*H with Axion Dark Matter Experiments}

For $\lambda_{H\sigma}<0$, the oscillations of the inflaton field after inflation allow for an efficient reheating. 
They result in the copious production of scalar field fluctuations via parametric resonance (``preheating''), thereby restoring the 
PQ symmetry after a few oscillations. The decay of the Higgs component of the inflaton into SM particles (mainly top quarks) reheats the Universe efficiently. 
The reheating temperature has been determined via 
lattice simulations of the evolution of the scalar fields in a an expanding Friedmann-Robertson-Walker background 
to be in the range 
$T_{\rm rh} 
\approx 10^{12-13}\,{\rm GeV}$~\cite{Ringwald:2022xif}. 
The predicted range is quite narrow, since the parameter space in SM*AS*H for the bosonic couplings of the PQ-field $\sigma$ is significantly constrained by requiring consistency of the inflationary predictions with CMB observations and of vacuum stability in the Higgs direction.  
Roughly, a given tensor-to-scalar ratio $r$ fixes $\xi_\sigma$ (see Fig.~\ref{fig:r_vs_ns}) which determines  the effective quartic coupling $\tilde\lambda_\sigma$  (see Fig.~\ref{fig:inflation_constraints} (left)). 
From the stability constraints it follows that $\tilde\lambda_{\sigma}$ cannot be very different from $\lambda_\sigma$. 
This means that choosing $r$ roughly specifies all the bosonic couplings of $\sigma$, which then settles the scalar field dynamics which determines preheating and reheating. 
For benchmark point BP1 in Fig.~\ref{fig:inflation_constraints}, 
the
simulations yielded a reheating temperature $T_{\rm rh}=9.7\times10^{12}$\,GeV, while for BP2 in the same figure, 
they resulted in  $T_{\rm rh}=2.0\times10^{12}$ GeV~\cite{Ringwald:2022xif}.

\begin{figure}[t]
\centerline{\includegraphics[width=0.85\textwidth]{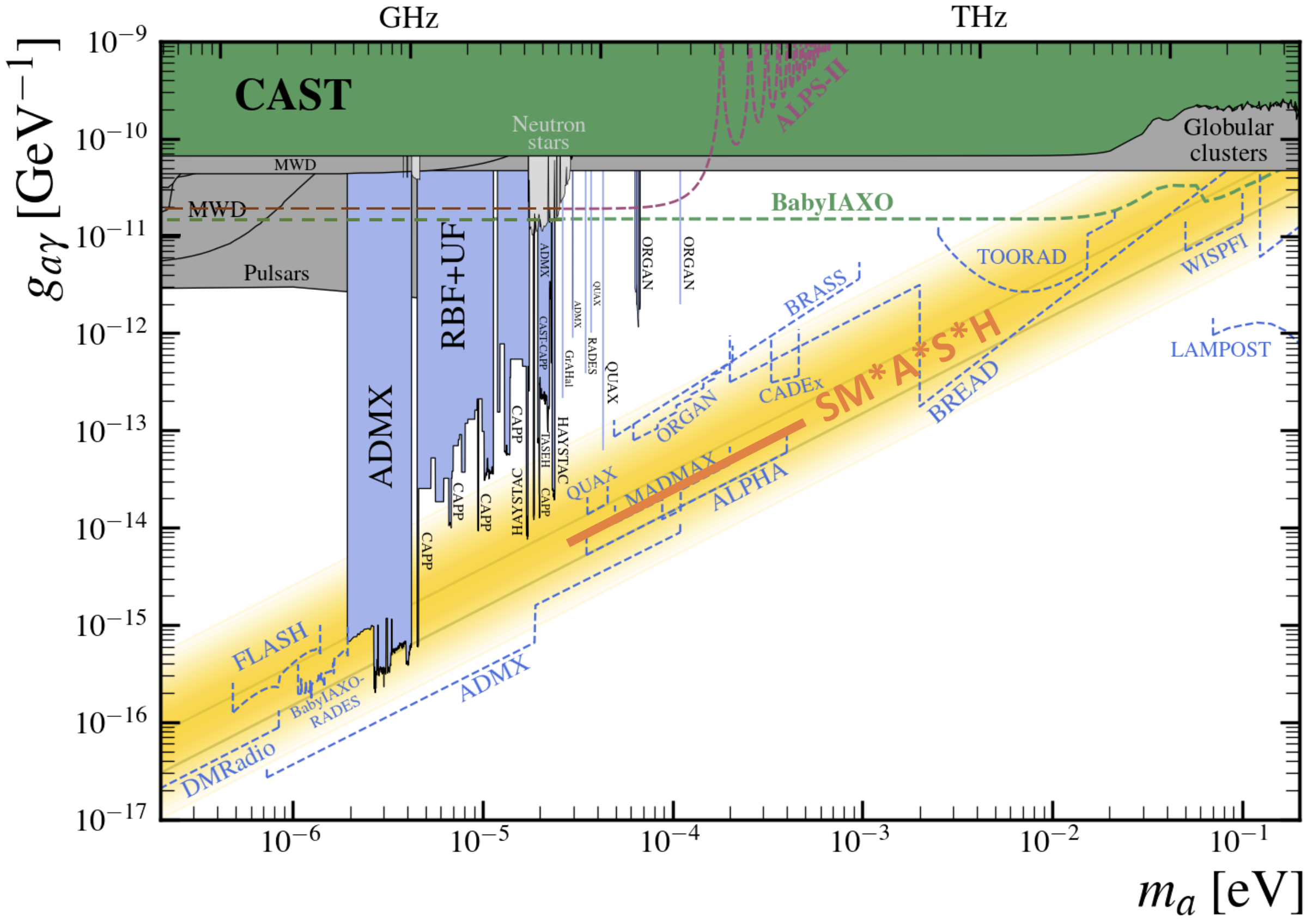}}
\caption[]{Electromagnetic coupling of the axion versus its mass. Haloscope exclusion regions (filled blue) and projected sensitivities (dashed blue lines) assume the axion to be 100\% of the halo dark matter. 
The solid orange band denotes the parameter region where the axion in SM*A*S*H is predicted to constitute 100\% of the cold dark matter
in the Universe.
Figure adapted from Ref.~\cite{AxionLimits}. 
}
\label{fig:photon_coupling_vs_mass}
\end{figure}

The axion is ``born'' when the PQ symmetry, which was restored during preheating, is broken again during the 
radiation dominated hot 
phase after reheating. 
The PQ phase transition is predicted to be second order, with critical 
temperature
$
T_c \simeq {2\sqrt{6\lambda_{\sigma}}v_\sigma}/{\sqrt{8(\lambda_{\sigma}+\lambda_{H\sigma})+\sum_iY_{ii}^2 + 6y^2}}
\simeq \lambda_\sigma^{1/4} v_\sigma \simeq  (2-8)\times 10^{-3}       \,v_\sigma$~\cite{Ballesteros:2016xej},
where 
the last two estimates   
take into account the parameter constraints arising from the requirements of vacuum stability and consistency of the inflationary 
predictions with CMB observations, respectively. 
After the PQ phase transition, the Universe consists of many causally disconnected patches, with sizes of order the Hubble radius, in which the axion field $\theta =a/f_a$ relaxes quickly to a uniform value between $-\pi$ and $\pi$. 
At the boundaries of these patches, one-dimensional topological defects - cosmic axion strings - are formed. 
The corresponding string loop network evolves during the Universe's expansion by tightening, oscillating, reconnecting, and collapsing, thereby continuously producing a population of effectively massless axions. 
This continues until the temperature of the Universe drops to around a GeV, when the Hubble expansion rate drops below the 
axion mass.  
Thereafter, axions are also produced from the decay of two-dimensional topological defects - axion domain walls - 
and from the realignment mechanism: 
within any Hubble size patch, the axion field starts to oscillate around the minimum of its potential, behaving as a cold dark matter fluid with equation of state 
$w_a=p_a/\rho_p\approx 0$. 

A conservative upper bound on the PQ symmetry breaking scale,  $f_a = v_{\sigma} < 2.3\times 10^{11}\, {\rm GeV}$, can be set by requiring that the predicted contribution to the dark matter abundance from the realignment mechanism alone does not exceed the observed dark matter abundance. This translates directly into a lower bound on the axion  
mass, $m_a \simeq 5.9\ {\rm \mu eV} \left( \frac{10^{12}\,{\rm GeV}}{f_a}\right)> 26\ \mu{\rm eV}$~\cite{Borsanyi:2016ksw}.  
Comparing with the sensitivity prospects of axion dark matter experiments (axion haloscopes) shown in Fig.~\ref{fig:photon_coupling_vs_mass}, we conclude:
{\bf ADMX~\cite{ADMX:2019uok}, BabyIAXO-RADES~\cite{Ahyoune:2023gfw}, CAPP~\cite{Yi:2022fmn}, FLASH~\cite{Alesini:2023qed}, or DMRadio~\cite{DMRadio:2022pkf} have good prospects to smash SM*A*S*H by discovering an axion with a mass below 26 micro-eV!} 

Unfortunately, the prediction of the axion dark matter contribution from cosmic strings and domain walls   
suffers from significant theoretical uncertainties 
which arise from the difficulty in quantifying precisely the energy loss processes of topological defects and the generated axion
spectra. In fact, the results 
from present state-of-the-art 
first principle classical field theory simulations~\cite{Kawasaki:2014sqa,Klaer:2017ond,Gorghetto:2018myk,Buschmann:2019icd,Hindmarsh:2019csc,Gorghetto:2020qws,Buschmann:2021sdq} 
still allow for two extreme possibilities: the
contribution from the decay of topological defects may either be negligible~\cite{Klaer:2017ond} or overwhelming~\cite{Gorghetto:2020qws} in comparison to the one from the realignment mechanism. 
In the latter case, a conservative lower bound on $f_a = v_{\sigma}\gtrsim 1.1\times 10^{10}\,{\rm GeV}$, corresponding to an upper bound on 
$m_a \lesssim 0.5\ {\rm meV}$,  
is obtained from the requirement that the produced axions constitute 100\% of the cold dark matter in the Universe. 
We infer from Fig.~\ref{fig:photon_coupling_vs_mass} that 
{\bf ALPHA~\cite{Lawson:2019brd} and MADMAX~\cite{Beurthey:2020yuq} have good prospects to detect the axion from SM*A*S*H if the latter constitutes 100\% of the dark matter in the galactic halo.\footnote{Large density variations in the initial state of the axion field after the PQ phase transition may lead to the formation of compact dark matter objects known as “miniclusters”. This leads to an increased theoretical uncertainty in the local axion density for dark matter detection in the laboratory. 
}}

\section{Searching for SM*A*S*H Signatures in the  Cosmic GW Background}

In this section, we are looking into the prospects to scrutinise SM*A*S*H in the far future, say in the second half of this century. 

\begin{figure}[h]
\begin{center}
\includegraphics[width=.95\textwidth]{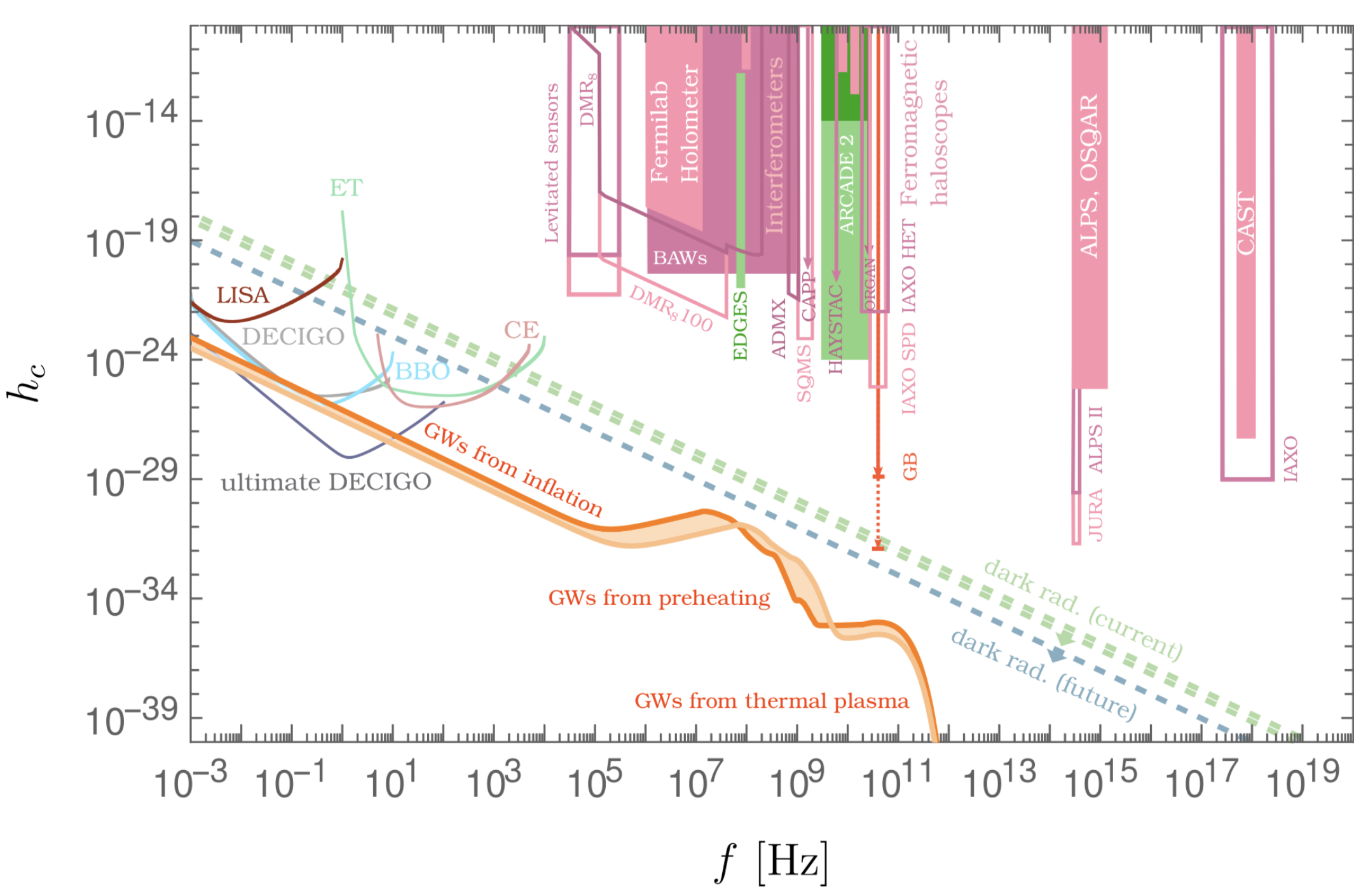}
\end{center}
\vskip-.5cm 
\caption{ Characteristic amplitude of primordial GWs in SM*A*S*H (orange; upper (lower) curve corresponding to the benchmark BP1 (BP2)  in Fig.~\ref{fig:r_vs_ns}) compared to present  (shaded areas) and projected sensitivities (colored solid lines)~\cite{LIGOScientific:2016wof,Seto:2001qf,Punturo:2010zz,LISA:2017pwj,BBO_proposal,Kuroyanagi:2014qza,Aggarwal:2020umq,Holometer:2016qoh,
Goryachev:2014yra,Akutsu:2008qv,Domcke:2020yzq,Ito:2019wcb,Ito:2020wxi,Ejlli:2019bqj,Domcke:2022rgu,
Ringwald:2020ist,Berlin:2021txa,Schmitz:Zenodo}. Indirect dark radiation constraints~\cite{Pagano:2015hma,Clarke:2020bil,Ghiglieri:2020mhm} are shown with dashed lines. 
Figure adapted from Ref.~\cite{Ringwald:2022xif}. 
}
\label{fig:hcOmega_smash_entire}
\end{figure}

Suppose, the CMB polarisation experiments have discovered, in the present decade, the B modes induced by 
the primordial tensor metric fluctuations originating from quantum fluctuations during inflation. 
This will result in a strong push towards the realisation of space-borne GW laser interferometers 
which are designed to search directly for the GWs from inflation, 
such as the Big Bang Observer (BBO)~\cite{BBO_proposal,Crowder:2005nr,Corbin:2005ny,Harry:2006fi} or the 
Deci-Hertz Interferometer Gravitational Wave Observatory 
(DECIGO)~\cite{Seto:2001qf,Kawamura:2006up}. 
In fact, in a frequency range around a Hz, these detectors have the prospected sensitivity to discover the stochastic GW background predicted from inflation in SM*A*S*H~\cite{Ringwald:2020vei}, as can be seen from Fig.~\ref{fig:hcOmega_smash_entire}. 
%
%

\begin{figure}[h]
\centering
\includegraphics[width=0.5\linewidth]{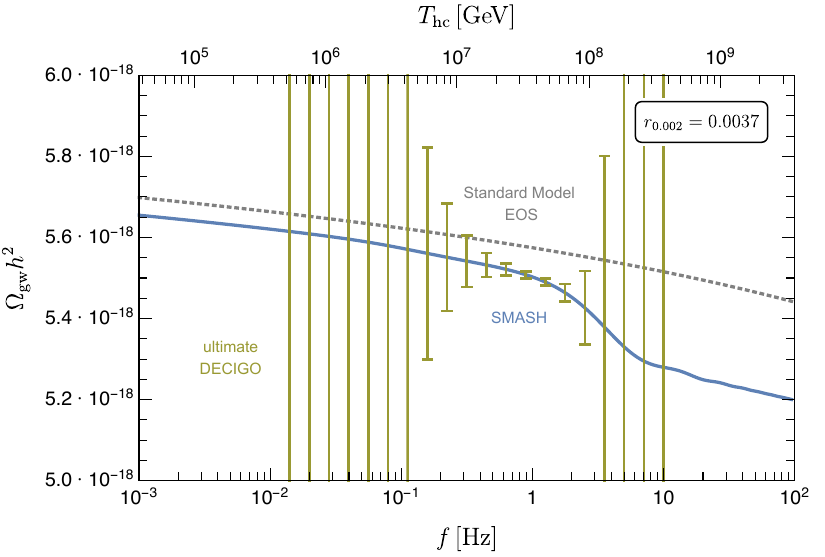}\includegraphics[width=0.45\textwidth]{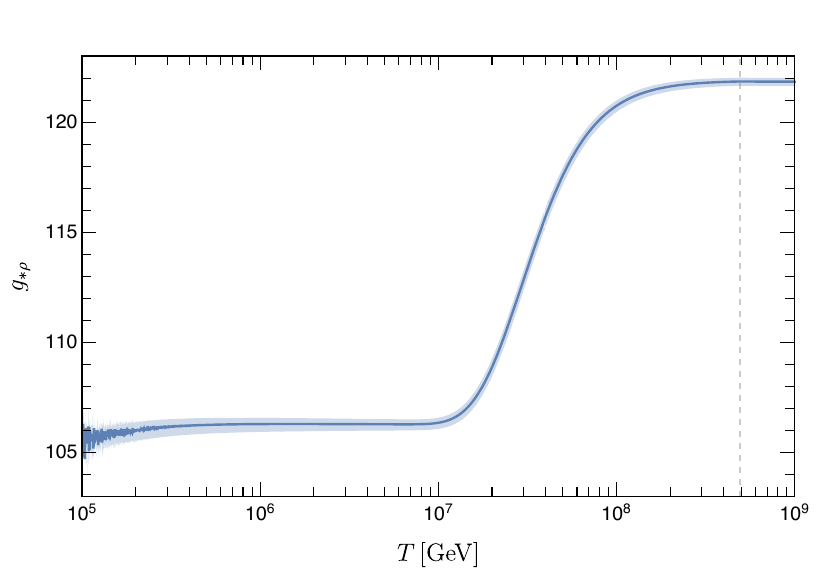}
\caption{
{\em Left panel:} 
The spectrum of primordial GWs predicted in SM*A*S*H for BP2 and $f_a=v_\sigma = 1.2\times 10^{11}\,\mathrm{GeV}$ (solid blue line),
corresponding to $T_c\simeq 5\times 10^9$\,GeV, 
in comparison with the results obtained by using the effective relativistic degrees of freedom in the SM.
Dark yellow bars correspond to the prospected sensitivity of ultimate DECIGO.
{\em Right panel:} 
Temperature dependence of $g_{*\rho}(T)$ for some specific choice of the parameters in SM*A*S*H (similar to BP1 with  
$f_a=v_\sigma = 1.2\times 10^{11}\,\mathrm{GeV}$) that leads to 
$T_c \simeq 5 \times 10^8$ GeV. 
Shaded regions represent the uncertainty due to the choice of the renormalization scale and estimate of the QCD corrections.
The gray dashed line corresponds to the critical temperature of the PQ phase transition. 
Both figures adapted from Ref.~\cite{Ringwald:2020vei}.
}
\label{fig:gwbroad_sensitivity}
\end{figure}

After a successful detection, one may envisage an upgrade of the detectors, such that their noise curves are determined by the quantum limit, for example to an ``ultimate'' DECIGO~\cite{Kuroyanagi:2014qza}. The latter will enable a precise measurement of the shape of the spectrum in this frequency range, see Fig.~\ref{fig:hcOmega_smash_entire}.  
Intriguingly, SM*A*S*H predicts a step in the spectrum within the sensitivity frequency window of ultimate DECIGO, 
see Fig.~\ref{fig:gwbroad_sensitivity} (left). 
This step is an observational signature of 
the second order PQ phase transition in SM*A*S*H, 
resulting in a drastic change in the equation of state of the thermal plasma, notably in  
the number of relativistic degrees of freedom, $g_{\ast \rho}(T)$, when the temperature drops below the critical 
temperature, $2\times 10^7\,\mathrm{GeV}\lesssim T_c  \simeq  
\lambda_{\sigma}^{1/4}\,v_\sigma 
\lesssim 2\times 10^9\,\mathrm{GeV}$,\footnote{Here, we have used for the range of $v_\sigma$ the one 
in which the axion can be 100\% of the cold dark matter},  see Fig.~\ref{fig:gwbroad_sensitivity} (right). 
Entropy conservation implies that the growth rate of the Hubble radius ${\mathcal H}^{-1}$ is diminished during the PQ 
phase transition.
Thus, the rate at which primordial tensor modes cross into the horizon is changed during the transition, resulting 
in a step in the spectrum at frequencies corresponding to the Hubble rate at the transition, 
$f = \frac{a(T_c) {\mathcal H}(T c)}{2\pi a_0}  \approx 
\,\mathrm{Hz}
\left(\frac{T_{c}}{10^8\,\mathrm{GeV}}\right)$.
%
%
Clearly, the PQ scale $v_\sigma$ is the crucial parameter determining the spectrum of the primordial GWs from inflation in SM*A*S*H,
since it affects $T_c\propto v_\sigma$  
and therefore changes the frequency range at which the step appears.
However, in second half of this century,  when ultimate DECIGO may operate, and in the case that the axion constitutes the main component of cold dark matter, 
axion haloscopes such as ALPHA and MADMAX should have detected the axion already and
determined its mass with high precision, from which one can infer its decay constant,  which is, in SM*A*S*H, equal to the PQ scale,
$f_a=v_\sigma$.   
Therefore, a detection of the step in the GW spectrum will yield further information about SM*A*S*H,
such as the quartic PQ field coupling $\lambda_\sigma$ and the Higgs portal coupling $\lambda_{H\sigma}$, which can be 
cross-correlated with the one obtained from CMB polarisation observations. 
In this sense, {\bf future GW observations of the GWs from inflation can be used to probe the details of the PQ sector that may not be reached in other high-energy experiments.}

But this is not the end of the story. SM*A*S*H has also definite predictions for stochastic GW backgrounds from other epochs of 
its cosmological history~\cite{Ringwald:2022xif}, notably from inflaton fragmentation during preheating and from the thermal plasma at the beginning 
of the hot thermal radiation-dominated stage, see Fig.~\ref{fig:hcOmega_smash_entire}. 
The contributions from the different epochs are not independent and their features are correlated:
each epoch determines the initial conditions for the subsequent one. 
{\bf A hypothetical detection of the complete spectrum in different frequency ranges would allow to cross-check for the correlations predicted in SM*A*S*H, opening new possibilities for falsifying the model.} 
For example, 
a hypothetical measurement of the GW spectrum between $0.1$ MHz and $100$ GHz could potentially determine the scale of inflaton fragmentation after inflation, related to the peak frequency of the GWs from preheating, and the  maximum temperature and the number of relativistic degrees of freedom of the hot Big Bang plasma, which fix the amplitude and peak of the GWs from the thermal 
plasma~\cite{Ghiglieri:2015nfa,Ghiglieri:2020mhm,Ringwald:2020ist}.  This could provide an unprecedented window into the physics of the very early universe. 
However, probing the GWs generated by preheating and the thermal plasma requires much progress in the detection of ultra high frequency GWs, as is obvious from Fig.~\ref{fig:hcOmega_smash_entire}. Fortunately, a worldwide initiative towards this goal has already started~\cite{Aggarwal:2020olq}.

\section{I Have a Dream}

Let us end this proceeding's contribution with a pipe dream of mine concerning the establishment of SM*A*S*H as the theory 
of particle physics and cosmology. 

In the early 2030's, the CMB polarisation experiments CMB-S4 and LiteBIRD 
discover B modes induced by inflationary GWs consistent with the prediction from SM*A*S*H in Fig.~\ref{fig:r_vs_ns}. 
Again in the 2030's, the axion haloscopes ALPHA and MADMAX discover axion dark matter in the mass region 
predicted by SM*A*S*H in Fig.~\ref{fig:photon_coupling_vs_mass}. In the 2060's, a GW laser interferometer with 
sensitivity in the Hz range similar
to BBO or DECIGO in Fig.~\ref{fig:hcOmega_smash_entire} is launched which directly detects the stochastic GW background 
predicted from inflation in SM*A*S*H shown in the same figure. In the 2080's, an upgrade of the latter space-borne GW detector
similar to ultimate DECIGO discovers the step in the inflationary GW background  in Fig.~\ref{fig:gwbroad_sensitivity} arising from the PQ phase transition in SM*A*S*H. Moreover, the SM*A*S*H parameter ranges inferred from the position of the step is consistent with the 
parameter ranges inferred from the measurements at CMB polarisation experiments and axion haloscopes. 
Of course, this motivates then to develop detectors also for ultra high frequency GWs in order to 
probe the SM*A*S*H predictions  in Fig.~\ref{fig:hcOmega_smash_entire} from preheating and the thermal plasma. However, 
I do not dare to give a time scale for this detection because this experimental field is still in its infancy.

\section*{Acknowledgments}
Special thanks to Guillermo Ballesteros, Kenichi Saikawa, and Carlos Tamarit  for valuable comments on the draft.
This work has been partially funded by the Deutsche Forschungsgemeinschaft (DFG, German Research Foundation) 
under Germany’s Excellence Strategy - EXC 2121 Quantum Universe - 390833306  and under 
- 491245950.
This article/publication is based upon work from COST Action COSMIC WISPers CA21106, supported by COST (European Cooperation in Science and Technology).


\begin{thebibliography}{99}

\bibitem{Ballesteros:2016euj}
G.~Ballesteros, J.~Redondo, A.~Ringwald and C.~Tamarit,
Phys. Rev. Lett. \textbf{118} (2017) no.7, 071802
[arXiv:1608.05414 [hep-ph]].

\bibitem{Ballesteros:2016xej}
G.~Ballesteros, J.~Redondo, A.~Ringwald and C.~Tamarit,
JCAP \textbf{08} (2017), 001
[arXiv:1610.01639 [hep-ph]].

\bibitem{Boucenna:2014uma}
S.~M.~Boucenna, S.~Morisi, Q.~Shafi and J.~W.~F.~Valle,
Phys. Rev. D \textbf{90} (2014) no.5, 055023
[arXiv:1404.3198 [hep-ph]].

\bibitem{Ballesteros:2019tvf}
G.~Ballesteros, J.~Redondo, A.~Ringwald and C.~Tamarit,
Front. Astron. Space Sci. \textbf{6} (2019), 55
[arXiv:1904.05594 [hep-ph]].

\bibitem{Sopov:2022bog}
A.~H.~Sopov and R.~R.~Volkas,
Phys. Dark Univ. \textbf{42} (2023), 101381
[arXiv:2206.11598 [hep-ph]].

\bibitem{Berbig:2022pye}
M.~Berbig,
JCAP \textbf{11} (2022), 042
[arXiv:2207.08142 [hep-ph]].

\bibitem{Dutta:2023lbw}
J.~Dutta, M.~Matlis, G.~Moortgat-Pick and A.~Ringwald,
[arXiv:2309.10857 [hep-ph]].

\bibitem{Degrassi:2012ry}
G.~Degrassi, S.~Di Vita, J.~Elias-Miro, J.~R.~Espinosa, G.~F.~Giudice, G.~Isidori and A.~Strumia,
JHEP \textbf{08} (2012), 098
[arXiv:1205.6497 [hep-ph]].

\bibitem{Lebedev:2012zw}
O.~Lebedev,
Eur. Phys. J. C \textbf{72} (2012), 2058
[arXiv:1203.0156 [hep-ph]].

\bibitem{Elias-Miro:2012eoi}
J.~Elias-Miro, J.~R.~Espinosa, G.~F.~Giudice, H.~M.~Lee and A.~Strumia,
JHEP \textbf{06} (2012), 031
[arXiv:1203.0237 [hep-ph]].

\bibitem{Ringwald:2022xif}
A.~Ringwald and C.~Tamarit,
Phys. Rev. D \textbf{106} (2022) no.6, 063027
[arXiv:2203.00621 [hep-ph]].

\bibitem{Peccei:1977hh}
R.~D.~Peccei and H.~R.~Quinn,
Phys. Rev. Lett. \textbf{38} (1977), 1440-1443

\bibitem{Weinberg:1977ma}
S.~Weinberg,
Phys. Rev. Lett. \textbf{40} (1978), 223-226

\bibitem{Wilczek:1977pj}
F.~Wilczek,
Phys. Rev. Lett. \textbf{40} (1978), 279-282

\bibitem{Preskill:1982cy}
J.~Preskill, M.~B.~Wise and F.~Wilczek,
Phys. Lett. B \textbf{120} (1983), 127-132

\bibitem{Abbott:1982af}
L.~F.~Abbott and P.~Sikivie,
Phys. Lett. B \textbf{120} (1983), 133-136

\bibitem{Dine:1982ah}
M.~Dine and W.~Fischler,
Phys. Lett. B \textbf{120} (1983), 137-141

\bibitem{Minkowski:1977sc}
P.~Minkowski,
Phys. Lett. B \textbf{67} (1977), 421-428

\bibitem{Gell-Mann:1979vob}
M.~Gell-Mann, P.~Ramond and R.~Slansky,
Conf. Proc. C \textbf{790927} (1979), 315-321
[arXiv:1306.4669 [hep-th]].

\bibitem{Yanagida:1979as}
T.~Yanagida,
Conf. Proc. C \textbf{7902131} (1979), 95-99
KEK-79-18-95.

\bibitem{Mohapatra:1979ia}
R.~N.~Mohapatra and G.~Senjanovic,
Phys. Rev. Lett. \textbf{44} (1980), 912

\bibitem{Fukugita:1986hr}
M.~Fukugita and T.~Yanagida,
Phys. Lett. B \textbf{174} (1986), 45-47

\bibitem{Spokoiny:1984bd}
B.~L.~Spokoiny,
Phys. Lett. B \textbf{147} (1984), 39-43

\bibitem{Futamase:1987ua}
T.~Futamase and K.~i.~Maeda,
Phys. Rev. D \textbf{39} (1989), 399-404

\bibitem{Salopek:1988qh}
D.~S.~Salopek, J.~R.~Bond and J.~M.~Bardeen,
Phys. Rev. D \textbf{40} (1989), 1753

\bibitem{Fakir:1990eg}
R.~Fakir and W.~G.~Unruh,
Phys. Rev. D \textbf{41} (1990), 1783-1791

\bibitem{Bezrukov:2007ep}
F.~L.~Bezrukov and M.~Shaposhnikov,
Phys. Lett. B \textbf{659} (2008), 703-706
[arXiv:0710.3755 [hep-th]].

\bibitem{Barbon:2009ya}
J.~L.~F.~Barbon and J.~R.~Espinosa,
Phys. Rev. D \textbf{79} (2009), 081302
[arXiv:0903.0355 [hep-ph]].

\bibitem{Burgess:2009ea}
C.~P.~Burgess, H.~M.~Lee and M.~Trott,
JHEP \textbf{09} (2009), 103
[arXiv:0902.4465 [hep-ph]].

\bibitem{BICEP:2021xfz}
P.~A.~R.~Ade \textit{et al.} [BICEP and Keck],
Phys. Rev. Lett. \textbf{127} (2021) no.15, 151301
[arXiv:2110.00483 [astro-ph.CO]].

\bibitem{SimonsObservatory:2018koc}
P.~Ade \textit{et al.} [Simons Observatory],
JCAP \textbf{02} (2019), 056
[arXiv:1808.07445 [astro-ph.CO]].

\bibitem{LiteBIRD:2022cnt}
E.~Allys \textit{et al.} [LiteBIRD],
PTEP \textbf{2023} (2023) no.4, 042F01
[arXiv:2202.02773 [astro-ph.IM]].

\bibitem{Abazajian:2019eic}
K.~Abazajian
\textit{et al.}, 
[arXiv:1907.04473 [astro-ph.IM]].

\bibitem{Borsanyi:2016ksw}
S.~Borsanyi
\textit{et al.},  
Nature \textbf{539} (2016) no.7627, 69-71
[arXiv:1606.07494 [hep-lat]].


\bibitem{AxionLimits}
C. O'Hare, 
doi:10.5281/zenodo.3932430,
https://github.com/cajohare/AxionLimits

\bibitem{ADMX:2019uok}
T.~Braine \textit{et al.} [ADMX],
Phys. Rev. Lett. \textbf{124} (2020) 
101303
[arXiv:1910.08638 [hep-ex]].

\bibitem{Ahyoune:2023gfw}
S.~Ahyoune
\textit{et al.}, 
Annalen Phys. \textbf{535} (2023) no.12, 2300326
[arXiv:2306.17243 [physics.ins-det]].

\bibitem{Yi:2022fmn}
A.~K.~Yi
\textit{et al.}, 
Phys. Rev. Lett. \textbf{130} (2023) no.7, 071002
[arXiv:2210.10961 [hep-ex]].

\bibitem{Alesini:2023qed}
D.~Alesini
\textit{et al.}, 
Phys. Dark Univ. \textbf{42} (2023), 101370
[arXiv:2309.00351 [physics.ins-det]].



\bibitem{DMRadio:2022pkf}
L.~Brouwer \textit{et al.} [DMRadio],
Phys. Rev. D \textbf{106} (2022) no.10, 103008
[arXiv:2204.13781 [hep-ex]].


\bibitem{Kawasaki:2014sqa}
M.~Kawasaki, K.~Saikawa and T.~Sekiguchi,
Phys. Rev. D \textbf{91} (2015) no.6, 065014
[arXiv:1412.0789 [hep-ph]].

\bibitem{Klaer:2017ond}
V.~B.~Klaer and G.~D.~Moore,
JCAP \textbf{11} (2017), 049
[arXiv:1708.07521 [hep-ph]].

\bibitem{Gorghetto:2018myk}
M.~Gorghetto, E.~Hardy and G.~Villadoro,
JHEP \textbf{07} (2018), 151
[arXiv:1806.04677 [hep-ph]].

\bibitem{Buschmann:2019icd}
M.~Buschmann, J.~W.~Foster and B.~R.~Safdi,
Phys. Rev. Lett. \textbf{124} (2020) no.16, 161103
[arXiv:1906.00967 [astro-ph.CO]].

\bibitem{Hindmarsh:2019csc}
M.~Hindmarsh, J.~Lizarraga, A.~Lopez-Eiguren and J.~Urrestilla,
Phys. Rev. Lett. \textbf{124} (2020) no.2, 021301
[arXiv:1908.03522 [astro-ph.CO]].

\bibitem{Gorghetto:2020qws}
M.~Gorghetto, E.~Hardy and G.~Villadoro,
SciPost Phys. \textbf{10} (2021) no.2, 050
[arXiv:2007.04990 [hep-ph]].

\bibitem{Buschmann:2021sdq}
M.~Buschmann, J.~W.~Foster, A.~Hook, A.~Peterson, D.~E.~Willcox, W.~Zhang and B.~R.~Safdi,
Nature Commun. \textbf{13} (2022) no.1, 1049
[arXiv:2108.05368 [hep-ph]].

\bibitem{Lawson:2019brd}
M.~Lawson, A.~J.~Millar, M.~Pancaldi, E.~Vitagliano and F.~Wilczek,
Phys. Rev. Lett. \textbf{123} (2019) no.14, 141802
[arXiv:1904.11872 [hep-ph]].

\bibitem{Beurthey:2020yuq}
S.~Beurthey, N.~B\"ohmer, P.~Brun, A.~Caldwell, L.~Chevalier, C.~Diaconu, G.~Dvali, P.~Freire, E.~Garutti and C.~Gooch, \textit{et al.}
[arXiv:2003.10894 [physics.ins-det]].

\bibitem{BBO_proposal} 
S.~Phinney {\it et al.}, ``The Big Bang Observer: Direct detection
of gravitational waves from the birth of the Universe to the Present'', NASA Mission
Concept Study (2004).

\bibitem{Crowder:2005nr}
  J.~Crowder and N.~J.~Cornish,
  Phys.\ Rev.\ D {\bf 72} (2005) 083005
  [gr-qc/0506015].


\bibitem{Corbin:2005ny}
  V.~Corbin and N.~J.~Cornish,
  Class.\ Quant.\ Grav.\  {\bf 23} (2006) 2435
  [gr-qc/0512039].


\bibitem{Harry:2006fi}
  G.~M.~Harry, P.~Fritschel, D.~A.~Shaddock, W.~Folkner and E.~S.~Phinney,
  Class.\ Quant.\ Grav.\  {\bf 23} (2006) 4887
   Erratum: [Class.\ Quant.\ Grav.\  {\bf 23} (2006) 7361].


\bibitem{Seto:2001qf}
N.~Seto, S.~Kawamura and T.~Nakamura,
Phys. Rev. Lett. \textbf{87} (2001), 221103
[arXiv:astro-ph/0108011 [astro-ph]].



\bibitem{Kawamura:2006up}
  S.~Kawamura {\it et al.},
  Class.\ Quant.\ Grav.\  {\bf 23} (2006) S125.


\bibitem{Ringwald:2020vei}
A.~Ringwald, K.~Saikawa and C.~Tamarit,
JCAP \textbf{02} (2021), 046
[arXiv:2009.02050 [hep-ph]].

\bibitem{LIGOScientific:2016wof}
B.~P.~Abbott \textit{et al.} [LIGO Scientific],
Class. Quant. Grav. \textbf{34} (2017) no.4, 044001
[arXiv:1607.08697 [astro-ph.IM]].


\bibitem{Punturo:2010zz}
M.~Punturo
\textit{et al.}, 
Class. Quant. Grav. \textbf{27} (2010), 194002

\bibitem{LISA:2017pwj}
P.~Amaro-Seoane \textit{et al.} [LISA],
[arXiv:1702.00786 [astro-ph.IM]].

\bibitem{Kuroyanagi:2014qza}
S.~Kuroyanagi, K.~Nakayama and J.~Yokoyama,
PTEP \textbf{2015} (2015) no.1, 013E02
[arXiv:1410.6618 [astro-ph.CO]].

\bibitem{Aggarwal:2020umq}
N.~Aggarwal, G.~P.~Winstone, M.~Teo, M.~Baryakhtar, S.~L.~Larson, V.~Kalogera and A.~A.~Geraci,
Phys. Rev. Lett. \textbf{128} (2022) no.11, 111101
[arXiv:2010.13157 [gr-qc]].

\bibitem{Holometer:2016qoh}
A.~S.~Chou \textit{et al.} [Holometer],
Phys. Rev. D \textbf{95} (2017) no.6, 063002
[arXiv:1611.05560 [astro-ph.IM]].

\bibitem{Goryachev:2014yra}
M.~Goryachev and M.~E.~Tobar,
Phys. Rev. D \textbf{90} (2014) no.10, 102005
[erratum: Phys. Rev. D \textbf{108} (2023) no.12, 129901]
[arXiv:1410.2334 [gr-qc]].

\bibitem{Akutsu:2008qv}
T.~Akutsu
 \textit{et al.},
Phys. Rev. Lett. \textbf{101} (2008), 101101
[arXiv:0803.4094 [gr-qc]].

\bibitem{Domcke:2020yzq}
V.~Domcke and C.~Garcia-Cely,
Phys. Rev. Lett. \textbf{126} (2021) no.2, 021104
[arXiv:2006.01161 [astro-ph.CO]].

\bibitem{Ito:2019wcb}
A.~Ito, T.~Ikeda, K.~Miuchi and J.~Soda,
Eur. Phys. J. C \textbf{80} (2020) no.3, 179
[arXiv:1903.04843 [gr-qc]].

\bibitem{Ito:2020wxi}
A.~Ito and J.~Soda,
Eur. Phys. J. C \textbf{80} (2020) no.6, 545
[arXiv:2004.04646 [gr-qc]].

\bibitem{Ejlli:2019bqj}
A.~Ejlli, D.~Ejlli, A.~M.~Cruise, G.~Pisano and H.~Grote,
Eur. Phys. J. C \textbf{79} (2019) no.12, 1032
[arXiv:1908.00232 [gr-qc]].

\bibitem{Domcke:2022rgu}
V.~Domcke, C.~Garcia-Cely and N.~L.~Rodd,
Phys. Rev. Lett. \textbf{129} (2022) no.4, 041101
[arXiv:2202.00695 [hep-ph]].

\bibitem{Ringwald:2020ist}
A.~Ringwald, J.~Sch\"utte-Engel and C.~Tamarit,
JCAP \textbf{03} (2021), 054
[arXiv:2011.04731 [hep-ph]].

\bibitem{Berlin:2021txa}
A.~Berlin, D.~Blas, R.~Tito D'Agnolo, S.~A.~R.~Ellis, R.~Harnik, Y.~Kahn and J.~Sch\"utte-Engel,
Phys. Rev. D \textbf{105} (2022) no.11, 116011
[arXiv:2112.11465 [hep-ph]].

\bibitem{Schmitz:Zenodo}
K.~Schmitz,
``New Sensitivity Curves for Gravitational-Wave Experiments'' [Data set], Zenodo, [http://doi.org/10.5281/zenodo.3689582].

\bibitem{Pagano:2015hma}
L.~Pagano, L.~Salvati and A.~Melchiorri,
Phys. Lett. B \textbf{760} (2016), 823-825
[arXiv:1508.02393 [astro-ph.CO]].

\bibitem{Clarke:2020bil}
T.~J.~Clarke, E.~J.~Copeland and A.~Moss,
JCAP \textbf{10} (2020), 002
[arXiv:2004.11396 [astro-ph.CO]].

\bibitem{Ghiglieri:2020mhm}
J.~Ghiglieri, G.~Jackson, M.~Laine and Y.~Zhu,
JHEP \textbf{07} (2020), 092
[arXiv:2004.11392 [hep-ph]].



\bibitem{Ghiglieri:2015nfa}
J.~Ghiglieri and M.~Laine,
JCAP \textbf{07} (2015), 022
[arXiv:1504.02569 [hep-ph]].

\bibitem{Aggarwal:2020olq}
N.~Aggarwal
 \textit{et al.}, 
Living Rev. Rel. \textbf{24} (2021) no.1, 4
[arXiv:2011.12414 [gr-qc]].



\end{thebibliography}
\end{document}